\documentclass[twocolumn,useAMS,usenatbib]{mnras} \usepackage{natbib}

\usepackage{amsmath,latexsym,amssymb} \usepackage{epsfig,graphicx} \topmargin-1cm
\usepackage{epstopdf}
\usepackage{float}
\usepackage{fixltx2e}
\usepackage{color}
\usepackage{url}
\graphicspath{{plots_figs/}}
\DeclareGraphicsExtensions{.pdf,.png,.jpg,.eps}

\def\reff@jnl#1{{\rm#1\/}}

\def\aj{\reff@jnl{AJ}}                  
\def\araa{\reff@jnl{ARA\&A}}            
\def\apj{\reff@jnl{ApJ}}                        
\def\apjl{\reff@jnl{ApJ}}               
\def\apjs{\reff@jnl{ApJS}}              
\def\apss{\reff@jnl{Ap\&SS}}            
\def\aap{\reff@jnl{A\&A}}               
\def\aapr{\reff@jnl{A\&A~Rev.}}         
\def\aaps{\reff@jnl{A\&AS}}             
\def\baas{\reff@jnl{BAAS}}              
\def\jcap{\reff@jnl{JCAP}}              
\def\jrasc{\reff@jnl{JRASC}}            
\def\memras{\reff@jnl{MmRAS}}           
\def\mnras{\reff@jnl{MNRAS}}            
\def\physrep{\reff@jnl{Phys.Rep.}}
\def\pra{\reff@jnl{Phys.Rev.A}}         
\def\prb{\reff@jnl{Phys.Rev.B}}         
\def\prc{\reff@jnl{Phys.Rev.C}}         
\def\prd{\reff@jnl{Phys.Rev.D}}         
\def\prl{\reff@jnl{Phys.Rev.Lett}}      
\def\pasp{\reff@jnl{PASP}}              
\def\pasj{\reff@jnl{PASJ}}              
\def\skytel{\reff@jnl{S\&T}}            
\def\solphys{\reff@jnl{Solar~Phys.}}    
\def\sovast{\reff@jnl{Soviet~Ast.}}     
\def\ssr{\reff@jnl{Space~Sci.Rev.}}     
\def\nat{\reff@jnl{Nature}}             

\newcommand{\beq}{\begin{equation}}
\newcommand{\eeq}{\end{equation}}
\newcommand{\beqa}{\begin{eqnarray}}
\newcommand{\eeqa}{\end{eqnarray}}

\title[BlueTides DMO : descendants of first quasars]{The descendants
  of the first quasars in the BlueTides simulation} \author[] {Ananth
  Tenneti$^1$\thanks{\tt vat@andrew.cmu.edu}, Tiziana Di
  Matteo$^1$\thanks{\tt tiziana@phys.cmu.edu}, Rupert Croft$^1$,
  ThomasJae Garcia$^1$, Yu Feng$^2$ \\$^1$McWilliams Center for
  Cosmology, Department of Physics, Carnegie Mellon University,
  Pittsburgh, PA 15213, USA \\$^2$ Berkeley Center for Cosmological Physics, Department of Physics, University of California Berkeley, Berkeley, CA 94720, USA}

\date{\today}
\begin{document}
\maketitle

\begin{abstract}
Supermassive blackholes with masses of a billion solar masses or more
are known to exist up to $z=7$. 
However,
the present-day environments of the descendants of first quasars is
not well understood and it is not known if they live in massive galaxy
clusters or more isolated galaxies at $z=0$. We use a dark matter-only
realization (BTMassTracer) of the BlueTides cosmological hydrodynamic simulation to
study the halo properties of the descendants of the most massive black
holes at $z=8$. We find that the descendants of the quasars with most
massive black holes are not amongst the most massive halos. They
reside in halos of with group-like ($\sim 10^{14}M_{\odot}$) masses, 
while the most
massive halos in the simulations are rich clusters with  masses $\sim 10^{15}
M_{\odot}$. The distribution of halo masses at low redshift is similar
to that of the descendants of least massive black holes, for a similar
range of halo masses at $z=8$, which indicates that they are likely to
exist in similar environments. 
By tracing back to the $z = 8$ progenitors of the most massive (cluster sized) halos at $z=0$; we find that their most likely black hole mass is less than $10^7 M_{\odot}$; they are clearly not amongst the most massive black holes. We also provide estimates for the likelihood of finding a high redshift quasar hosting a black hole with masses above $10^{7} M_{\odot}$ for a given halo mass at $z=0$. For halos above $10^{15} M_{\odot}$, there is only $20 \%$ probability that their $z=8$ progenitors hosted a blackhole with mass above $10^{7} M_{\odot}$.
\end{abstract}

\begin{keywords}
cosmology: early universe -- methods: numerical -- hydrodynamics -- galaxies: high-redshift -- 
quasars: supermassive black holes 
\end{keywords}

\section{Introduction} \label{S:intro}

The discovery of highly luminous quasars at high redshifts ($z>6$)
suggests that black holes of masses around 10 billion solar masses
existed in the early universe \citep{2015Natur.518..512W} and the
highest redshift quasar discovered so far is at $z\sim7$
\citep{2011Natur.474..616M}. However, the environments of the
descendants of the first quasars are yet to be understood and it is not
clear if they live in the most massive clusters of galaxies today or
in rather more isolated galaxies \citep[see e.g.,][]{2016Natur.532..340T}. In
\cite{2016Natur.532..340T}, the discovery of a black hole with mass
about 17 billion solar masses in an isolated galaxy was reported as
part of the MASSIVE survey \citep{2014ApJ...795..158M}, and which is likely to
be the descendant of a luminous quasar. In an earlier study
\citep{2011Natur.480..215M}, the hosts of the  massive black
holes at the present day were found to be in the centers of galaxy clusters.

In order to understand the properties of the descendants of quasars
numerically, we need hydrodynamic simulations of large computational
volume. In a previous study, \cite{2017MNRAS.467.4243D} used the
BlueTides hydrodynamic simulation \citep{2016MNRAS.455.2778F} to study
the origin of the most massive black holes at $z=8$.  BlueTides is the
largest-scale uniform volume cosmological hydrodynamic simulation with
sub-kpc resolution to date.  With the unprecedented combinations of
volume and resolution it is ideally suited for the study of the properties
of rare objects in the early Universe.

It has been found that the brightest quasars do not necessarily live
in the highest density regions (rather the strongest correlation was
found to be with low tidal field regions) of the early universe
consistent with the findings in
\cite{{2012ApJ...752...39T},{2013MNRAS.436..315F},{2017arXiv170406050U}},
suggesting that their descendants may not necessarily be among the
most massive halos at $z=0$. Consistent with this scenario,
\cite{2017arXiv170406050U} recently found using observations from
Hyper Suprime-Cam that luminous quasars do not live in the most
overdense regions at $z \sim 4$. \cite{2016MNRAS.460.1147B} used the
EAGLES cosmological hydrodynamic simulation to study the origin of
galaxies hosting super-massive black holes whose black hole masses at
$z=0$ are an order of magnitude more massive than that implied by
their bulge luminosities or masses, based on the $M_{BH}-M_{*}$
relation. These studies indicate that quasar descendants are most
likely not to live in privileged sites. However, we need to directly
track the descendants of the first quasars to $z=0$. It is not
computationally feasible to run the BlueTides hydrodynamic simulation
to $z=0$.  We instead use a dark matter-only simulation (BTMassTracer), evolved 
with the same initial conditions as the hydrodynamic simulation and match the
halos in both the simulations at $z=8$. By tracking the halo hosts
of the quasars to $z=0$ in the dark matter-only run, we can understand
the properties of the descendants of the first supermassive black
holes in BlueTides.

This paper is organized as follows. In Section~\ref{S:SimMethods}, we
discus simulations and methods. The results are presented in
Section~\ref{S:Results} and finally, we conclude in
Section~\ref{S:conclusions}.
             
\section{Simulations and methods}\label{S:SimMethods}

\begin{table}
\caption{\label{T:param}Simulation parameters: Box size (L$_\text{Box}$), initial redshift ($z_{i}$), final redshift ($z_{f}$), force softening length ($\epsilon _{grav}$), number of particles ($N_\text{Particle}$), mass of dark matter particle ($M_{\text{DM}}$) and mass of gas particle ($M_{\text{GAS}}$) }
\begin{tabular}{|c|c|c|}
\hline
 Parameters & Hydrodynamic & Dark Matter-Only \\ & (BT) & (BTMassTracer)\\
\hline
L$_\text{Box}$ ($h^{-1}$Mpc)& 400 & 400\\
$z_{i}$                   &  99  & 99 \\
$z_{f}$                   &  8   & 0 \\
$\epsilon _{grav}$ ($h^{-1}$kpc)& 1.5 & 7.4\\
$N_\text{Particle}$ & $2 \times 7040^{3}$ & $1760^{3}$\\
$M_{\text{DM}}$ ($h^{-1}M_{\odot}$) & $1.2 \times 10^{7}$ & $9.17 \times 10^{8}$\\
$M_{\text{GAS}}$($h^{-1}M_{\odot}$) & $2.36 \times 10^{6}$ & $0$\\
\hline
\end{tabular}
\end{table}

In this paper, we use the BlueTides (BT) MassTracer, a dark matter-only 
simulation, which is performed in a box of size, $400 h^{-1}Mpc$ on a
side. The simulation is performed with the MP-Gadget code using the
same initial conditions as the BlueTides hydrodynamic simulation
\citep{2016MNRAS.455.2778F} of galaxy formation, generated at
$z=99$. The BT has been evolved to $z=8$ with $2 \times 7040^{3}$ dark
matter and gas particles where the hydrodynamics is implemented with a
pressure-entropy formulation of Smoothed Particle
Hydrodynamics\citep{{2010MNRAS.405.1513R},{2013MNRAS.428.2840H}}. The
star formation is implemented based on a multi-phase star formation
model \citep{2003MNRAS.339..289S} and also including modifications
following \cite{2013MNRAS.436.3031V}. Gas is allowed to cool through
radiative processes \citep{1996ApJS..105...19K} and metal cooling
\cite{2014MNRAS.444.1518V}. The formation of molecular hydrogen and
its effect on star formation at low metallicities is modeled based on
the prescription in \cite{2011ApJ...729...36K}. A type II supernovae
wind feedback model \cite{2010MNRAS.406..208O} is included as well as
black hole growth and feedback from active galactic nuclei (AGN) using
the super-massive black hole model developed in
\cite{2005Natur.433..604D}. The dark matter-only simulation is
performed with a lower resolution using $1760^{3}$ dark matter
particles and evolved to $z=0$. Here, the mass of each dark matter particle is $9.17 \times 10^{8}h^{-1}M_{\odot}$ and the force softening length is $7.4 h^{-1} \mathrm{kpc}$. 
The cosmological parameters in both
the simulations are chosen according to WMAP9
\citep{2013ApJS..208...19H}. In Table~\ref{T:param}, we list the box size (L$_\text{box}$), initial redshift ($z_{i}$), final redshift ($z_{f}$), force
softening length ($\epsilon _{grav}$), total number of particles including
dark matter and gas ($N_\text{Particle}$), mass of dark matter particle
($M_{\text{DM}}$) and initial mass of gas particle ($M_{\text{GAS}}$) for
the BT and BTMassTracer simulations.

Since both the simulations are performed with the same initial
conditions, it is possible to match halos across the hydrodynamic and
dark matter-only simulations at $z=8$. For each halo in BT, we choose
the halo in the BTMassTracer with less than $50 \%$ difference in halo mass
and whose minimum potentials are closest to each other. To study the
descendants of quasars in BT, we identify the most bound particle in
the matched halo at $z=8$ as a tracer of the black hole host galaxy in
the corresponding hydrodynamic simulation. This particle (and its
halo) in BTMassTracer can be tracked across all the redshifts to $z=0$ based
on its unique particle ID. The halo in which the tracer particle
resides is the descendant of the hosts of quasars at $z=8$ and we can
study their properties using the dark matter-only simulation.

\section{Results}\label{S:Results}

\begin{figure*}
\begin{center}
\includegraphics[width=3.2in]{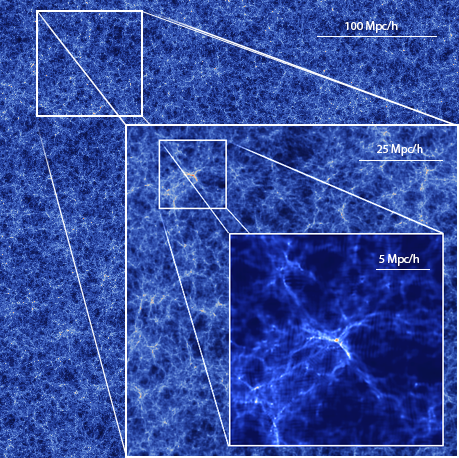}
\includegraphics[width=3.2in]{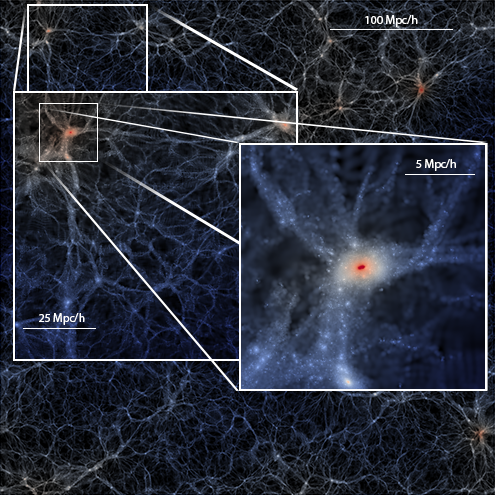}
\caption{\label{F:fig0a_6Mpc} Environment of the most massive black
  hole at $z=8$ in BT ({\it Left}) and its descendant at $z=0$ in BTMassTracer ({\it
    Right}).  The images show the dark matter density in a slice
through the entire simulation (width $400 h^{-1}Mpc$)
 and thickness 2 $h^{-1}Mpc$. In the left panel,
the intensity and color of the pixels is representative 
the density of dark matter
at each point. In the right panel, at lower redshift, the intensity
is proportional to the density and the color scale indicates
the gravitational potential.
}
\end{center}
\end{figure*}

\begin{figure*}
\begin{center}
\includegraphics[width=3.2in]{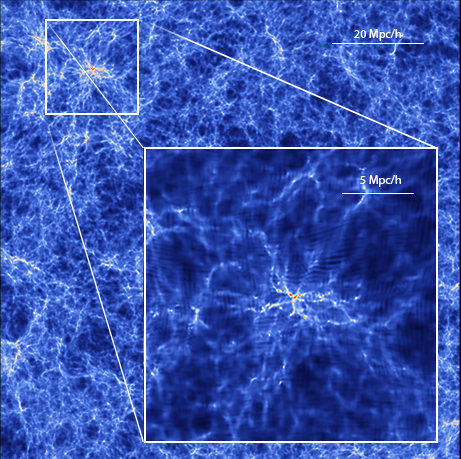}
\includegraphics[width=3.2in]{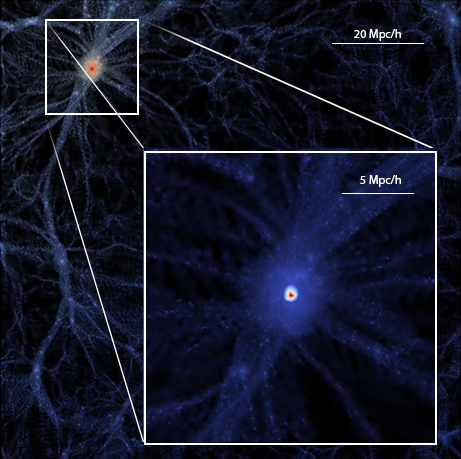}
\caption{\label{F:fig1a_6Mpc} Environment : Descendant of the $z=8$ most massive FOF halo in a $100 h^{-1}Mpc$ region at $z=8$ ({\it Left}) and $z=0$ ({\it Right}).}
\end{center}
\end{figure*} 
 
\begin{figure*}
\begin{center}
\includegraphics[width=3.2in]{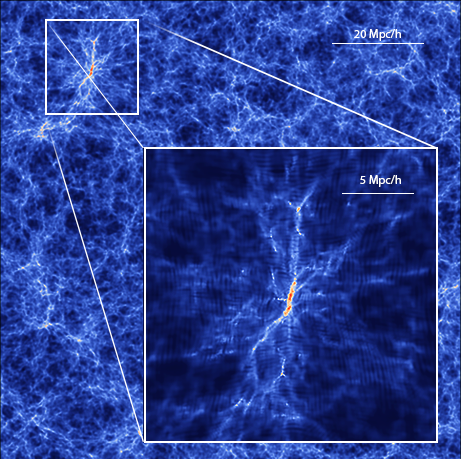}
\includegraphics[width=3.2in]{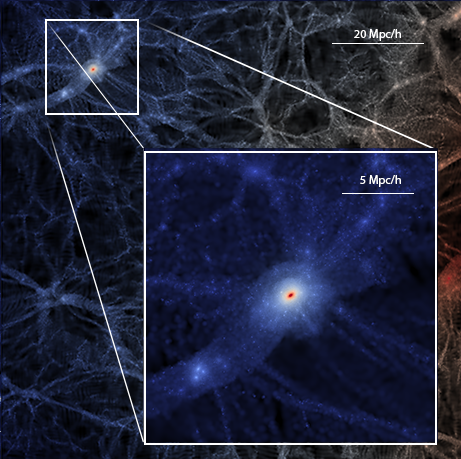}
\caption{\label{F:fig2a_6Mpc} Environment : $z=0$ most massive FOF halo in a $100 h^{-1}Mpc$ region at $z=8$ ({\it Left}) and $z=0$ ({\it Right}).}
\end{center}
\end{figure*} 

Before carrying out a statistical study of the descendants of high redshift 
quasars, we examine a few illustrative cases.
First, we track the tracer of the most massive blackhole at $z=8$ based on
the method described in Section~\ref{S:SimMethods} and identify the
host halo at $z=0$. Figure~\ref{F:fig0a_6Mpc} shows the distribution
of dark matter  in a slice through the simulation that
includes  the
massive black hole at $z=8$ and its tracer at $z=0$. The slices plotted are
2 ${\rm h^{-1}Mpc}$ thick and in each case we also show zooms into a 
100 ${\rm h^{-1}Mpc}$ wide and 20 ${\rm h^{-1}Mpc}$ wide  subvolume.
In order to  highlight the
variety of features present in the high and low density universe, we use
a color scale in the $z=0$ panel which varies with the gravitational
potential.
 We can see that, as expected, the 
density field is much more structured at redshift $z=0$ and visible filaments
extend to larger scales. The 
quasar descendant inhabits an elliptical halo with projected minor and
major axes with dimensions
$\sim 1 h^{-1}Mpc$ and $\sim 2 h^{-1}Mpc$. Its mass is $4.92 \times 10^{14} M_{\odot}$,
comparable to a large galaxy group such as the NGC 1600 group \citep{2016Natur.532..340T}.
 It is connected
to the surrounding large-scale structure by 5 obvious filaments at $z=0$,
and at $z=8$ it appears to lie at the intersection of 3 filaments. 
The extent of the ``protocluster'' at $z=8$ visible in 
the zoomed panel is not much larger
than the final cluster at $z=0$. This indicates that the 3 filament structures
seen at $z=8$ must have collapsed, virialised and consequently erased by $z=0$.
At $z=8$, the halo hosting the most massive black hole has a mass of $8.43 \times 10^{11} M_{\odot}$ and its descendant is the $242^{nd}$ most massive halo at $z=0$. 

We also trace the most massive halo at $z=8$ forward to $z=0$. This halo has
a mass of $2.77 \times 10^{12} M_{\odot}$ at $z=8$ and is about $3$ times more massive than the host of the most massive black hole. 
 Its environment at these two
redshifts is shown in Figure~\ref{F:fig1a_6Mpc}. The protocluster region
at $z=8$ in this case is more spherically symmetric, and less filamentary
than in Figure~\ref{F:fig0a_6Mpc}. These two objects are
extreme, and this finding is counter to the trend seen on average
by \cite{2017MNRAS.467.4243D} that early quasars preferentially occur in regions
with low tidal fields. At $z=0$ the halo that was the most massive at
$z=8$ is now the $258^{th}$  most massive with a mass
 of $4.79 \times 10^{14} M_{\odot}$, and appears more 
spherical than the quasar host descendant.

An alternative way to study these extreme objects is instead to track
halos backwards in time. We have done this in
Figure~\ref{F:fig2a_6Mpc}, where we show the most massive halo at
$z=0$ and environment around its progenitor at $z=8$. In this
case, the $z=0$ halo has a mass of $3.59 \times 10^{15} M_{\odot}$,
 comparable to the Coma
cluster.  At $z=8$, it lies in a $\sim 6 {\rm h^{-1}Mpc}$ filament,
and the progenitor halo has a mass $1.25 \times 10^{12} M_{\odot}$ 
 ( $\sim 45 \%$ of the mass of the most
massive halo at $z=8$).  Looking at the very top halos and black holes
in mass as we have done in these three figures, we can see that they
reside in obviously dense environments, and have similar halo masses.
It is not clear however what happens to the bulk of the population-
for example whether all early quasars have massive descendants, or
conversely all the massive halos at $z=0$ hosted massive black holes
at $z=8$. We can answer this by tracking a statistical sample of
objects.
      
\begin{figure}
\begin{center}
\includegraphics[width=3.2in]{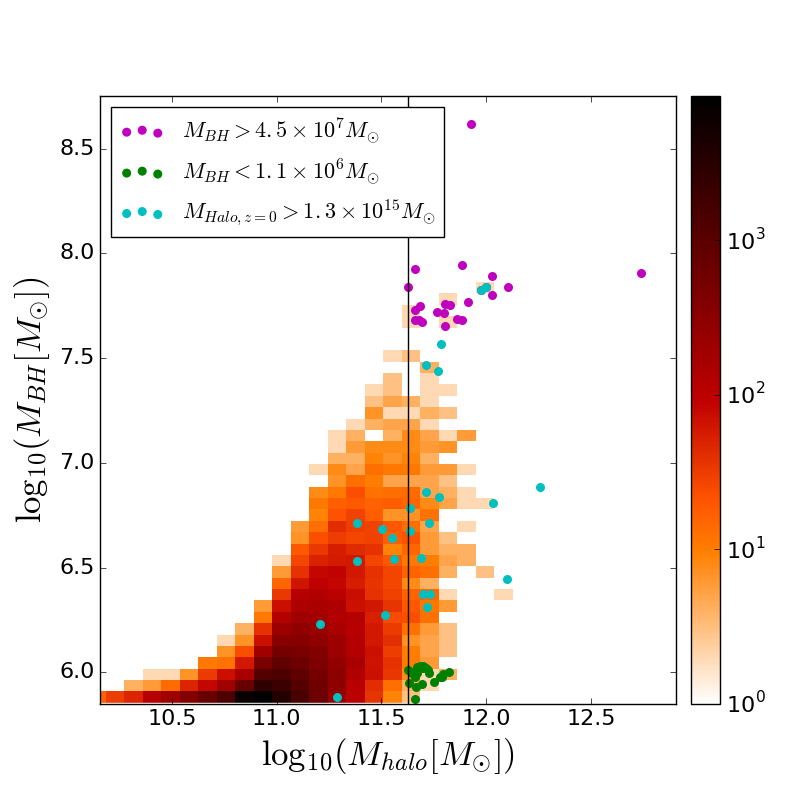}
\caption{\label{F:fig1_MbhMhalo} Two dimensional histogram of the
  logarithm of halo mass ($\log _{10} (M_{halo}[M_{\odot}])$) and the
  black hole mass ($\log _{10} (M_{BH}[M_{\odot}])$) in the BlueTides
  Hydrodynamic simulation at $z=8$. The 23 most massive black holes
  are shown by scatter points in purple, while the least massive
  blackholes are shown in green. We also show as cyan points, the
  blackholes in the progenitor halos of the most massive halos at
  $z=0$.  }
\end{center}
\end{figure}

In Figure~\ref{F:fig1_MbhMhalo}, we plot a two-dimensional histogram
of halo mass and the corresponding black hole mass at $z=8$ in the
BlueTides hydrodynamic simulation. The scatter points colored in 
magenta  in the figure show the halos with black hole mass, $M_{BH} > 4.5 \times 10^{7}M_{\odot}$ which correspond to the 23 most massive black holes at $z=8$.
We track these in the BlueTides MassTracer simulation. In a similar 
halo mass range, we also track an equal number of 
least massive
black holes (shown as green points) in order to compare the properties
of the descendants of the halos with massive quasars with those of
halos whose black hole mass is much lower, around that of the seed
black hole mass.  While matching halos in the BT hydrodynamic and 
BTMassTracer simulation, we note that the halo mass is different in the
matched halos due to baryonic effects and there can be differences,
which are mostly within $20 \%$.
 
\begin{figure*}
\begin{center}
\includegraphics[width=3.2in]{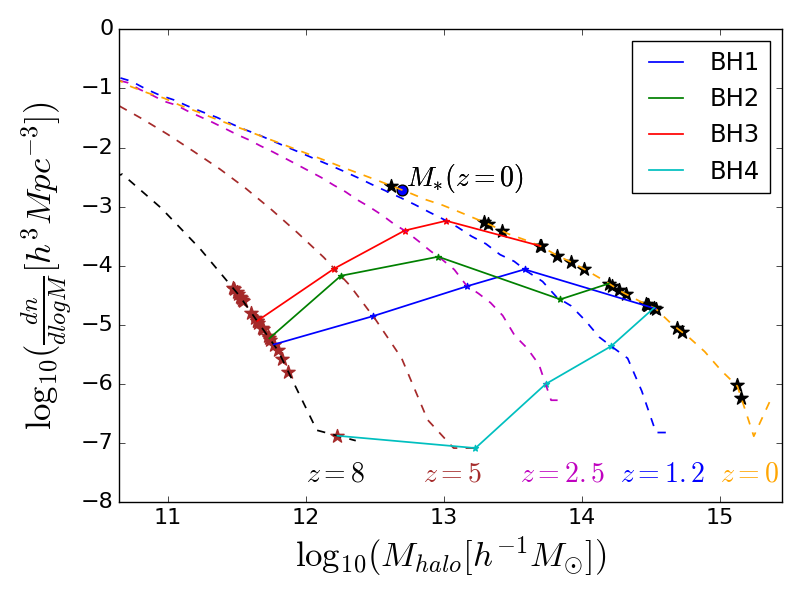}
\includegraphics[width=3.2in]{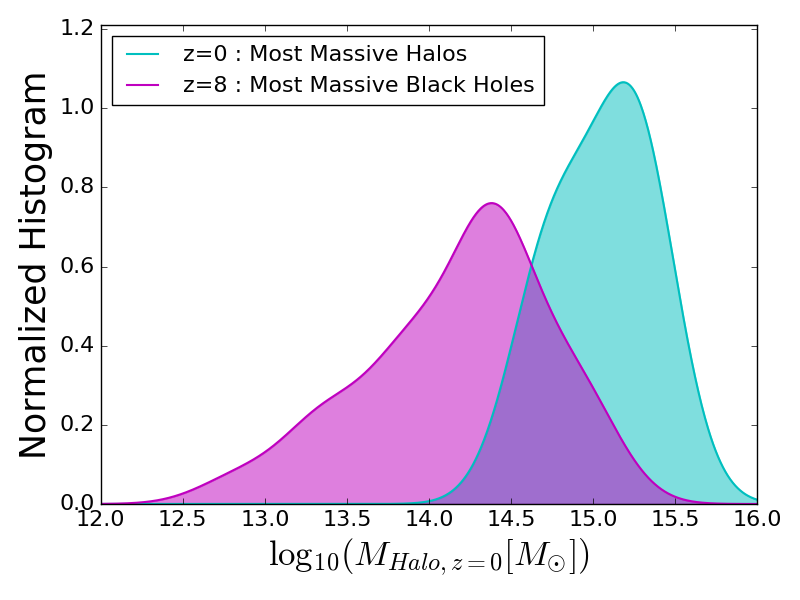}
\caption{\label{F:fig2_halomassfunc} {\it Left:} Halo Mass function at
  $z=8, 5, 2.5, 1.2, 0$ and the location of the tracked hosts of the 4
  most massive BH's on the mass function. {\it Right:} Normalized
  histogram of the halo masses at $z=0$ of the descendants of the most
  massive black holes at $z=8$ compared with the histogram of most
  massive halos at $z=0$. }
\end{center}
\end{figure*}

In the left panel of Figure~\ref{F:fig2_halomassfunc}, we plot the
halo mass function of the dark matter halos at $z=8, 5, 2.5, 1.2,
0$. We also plot as points on the halo mass function at $z=8$ and
$z=0$ the hosts of the most massive black hole tracer particles
(corresponding to the purple points in Figure~\ref{F:fig1_MbhMhalo}).
We find that at redshift $z=0$, the halo masses are distributed over a
wide range, compared to that at $z=8$. From the figure, we can
therefore clearly see that the descendants of the most massive black
holes at $z=8$ are not all the most massive halos at $z=0$. However,
the descendants are still relatively massive halos with mass above
$M_{*}= 5.0 \times 10^{12} h^{-1}M_{\odot}$ at $z=0$. The individual
tracks for the halo masses of the four most massive black hole tracers
are also shown in the figure. We can see that these objects
approximately keep their rank order over all redshifts.

To further understand the distribution of the halo masses, we plot the
normalized histogram of the host halo masses of the descendants of the
$z=8$ massive black holes, compared with the 23 most massive halos at
$z=0$ in the right panel of Figure~\ref{F:fig2_halomassfunc}. While
the most massive halos are distributed above $3.2 \times
10^{14}h^{-1}M_{\odot}$ with a peak in the distribution at $\sim 1.8
\times 10^{15}h^{-1}M_{\odot}$, the quasar descendants have a peak
distribution in the mass at $\sim 2.24 \times
10^{14}h^{-1}M_{\odot}$. These 23 most massive halos and 23 quasar
descendants have a space density of $4\times 10^{-7} {\rm h^3
  Mpc^{-3}}$. The massive halos are all Coma-like objects, whereas the
quasar descendants have masses an order of magnitude smaller,
comparable to galaxy groups with a space density $\sim 10^{-5} {\rm
  h^3 Mpc^{-3}}$.

\begin{figure*}
\begin{center}
\includegraphics[width=3.2in]{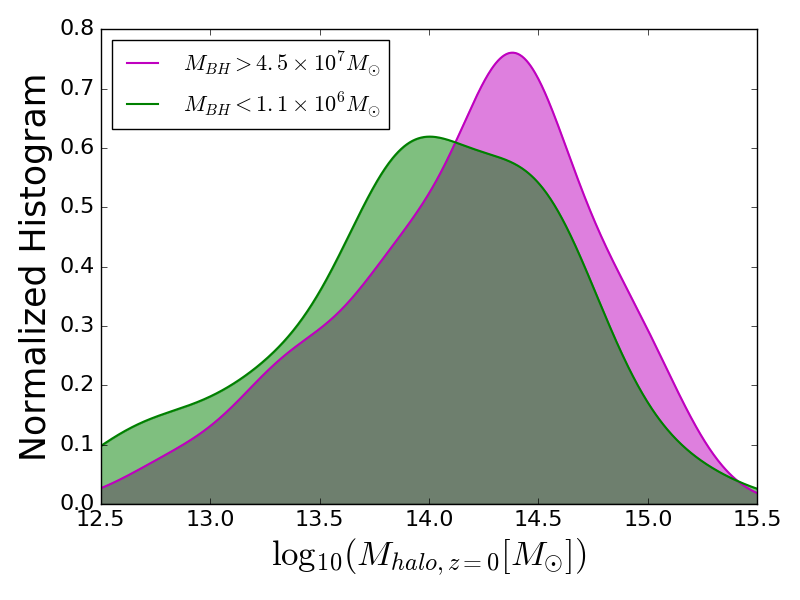}
\includegraphics[width=3.2in]{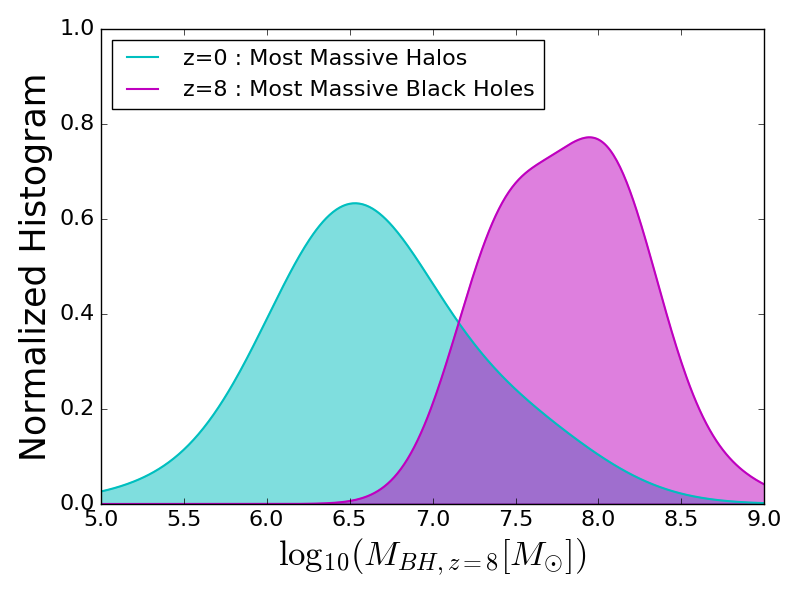}
\caption{\label{F:fig_Mhaloz0Mbh}{\it Left:} Normalized histogram of
  the halo masses at $z=0$ of the descendants of the most massive
  black holes ($M_{BH} > 10^{7.653}M_{\odot}$)and the least massive
  black holes ($M_{BH} < 10^{6.031}M_{\odot}$) at $z=8$. Note :
  KS-test statistic shows that the distributions are similar with
  $\alpha =0.26$ and $p$-value = $0.36$. {\it Right:} Normalized
  histogram of the black hole masses of the most massive black holes
  at $z=8$ and the black holes whose descendants are the most massive
  halos at $z=0$.}
\end{center}
\end{figure*}

We can similarly compare the distribution of halo masses for the
descendants of the least massive black holes at $z=8$ with those of
the most massive ones, whose halo masses at $z=8$ are in the similar
mass range. These are the green points in
Figure~\ref{F:fig1_MbhMhalo}.  Tracking forwards to $z=0$, we show the
normalized histogram of the halo masses for the descendants of the
least massive and the most massive black holes in the left panel of
Figure~\ref{F:fig_Mhaloz0Mbh}. We can clearly see that the
distribution of halo masses are in similar range in both cases. Based
on a KS-test, we have verified that the distributions are
statistically consistent with each other (see numerical values in the
figure caption). This is the case even though the black hole masses at
$z=8$ in the two subsamples are different by about two orders of
magnitude.

We have seen that after controlling for the effect of halo mass, high
redshift quasars do not have overly massive descendants. An
interesting related question is whether the most massive halos at
$z=0$ were hosting overly massive black holes at high redshift.  In
order to answer this, we need to find the $z=8$ black hole masses (in
the BlueTides run) of the most massive halos (in the BTMassTracer run)
at $z=0$.  First, we identify the tracers of the black hole particles
at $z=8$ which are in the 23
most massive halos at
$z=0$. This is done by identifying halos such that the most bound
particle in these halos at $z=8$ is present in the corresponding most
massive halos at $z=0$. By matching these halos in the BTMassTracer
 simulation at $z=8$ with that of the hydrodynamic
simulation, we can identify the black hole masses at $z=8$, whose
descendants are the most massive halos at $z=0$.

In Figure~\ref{F:fig1_MbhMhalo}, we show these black hole masses as
cyan points. We can see that they are spread throughout the range of
masses plotted, with most lying in between the most massive ($\sim
10^8 M_{\odot}$) and the least massive $\sim 10^{6} M_{\odot}$.  The
normalized histograms of these black hole masses are plotted in the
right panel of Figure~\ref{F:fig_Mhaloz0Mbh}, compared with the
distribution of the most massive black holes. From the figure, we can
see that the most massive halos at $z=0$ are the descendants of the
black holes mostly with masses less than $10^{7}M_{\odot}$ and the
peak in their mass distribution is at $\sim 3.2 \times 10^{6}
M_{\odot}$.  The most massive black holes have masses above $10^{7}
M_{\odot}$ with a peak in the distribution at $\sim 10^{8}
M_{\odot}$. This result is consistent with our discussion earlier that
the descendants of the most massive black holes at $z=8$ are not among
the the most massive halos.

\begin{figure}
\begin{center}
\includegraphics[width=3.2in]{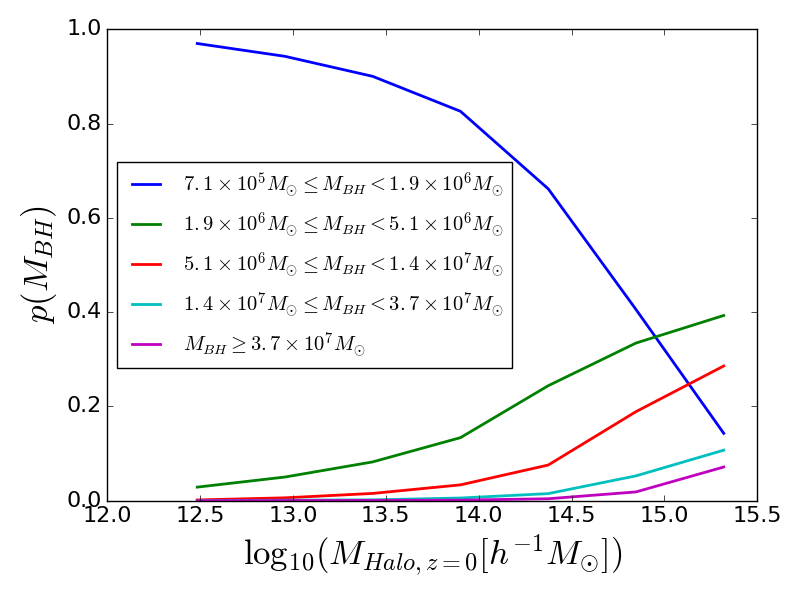}
\caption{\label{F:fig_Mbh_prob} Likelihood of the black hole mass of
  the halo at $z=8$, $M_{BH}$ is with in the range, $7.1 \times 10^{5}
  M_{\odot} \leq M_{BH} < 1.9 \times 10^{6} M_{\odot}$, $1.9 \times
  10^{6} M_{\odot} \leq M_{BH} < 5.1 \times 10^{6} M_{\odot}$, $5.1
  \times 10^{6} M_{\odot} \leq M_{BH} < 1.4 \times 10^{7} M_{\odot}$,
  $1.4 \times 10^{7} M_{\odot} \leq M_{BH} < 3.7 \times 10^{7}
  M_{\odot}$ or $ \geq 3.7 \times 10^{7} M_{\odot}$ given the halo
  mass of the descendant at $z=0$.}
\end{center}
\end{figure}

Finally, for a given halo mass at $z=0$, we can make predictions about
its black hole mass at $z=8$ from our analysis. In
Figure~\ref{F:fig_Mbh_prob}, we plot the probability that the black
hole mass at $z=8$ falls within certain mass bins. We
choose $7.1 \times 10^{5}
M_{\odot} \leq M_{BH} < 1.9 \times 10^{6} M_{\odot}$, $1.9 \times
10^{6} M_{\odot} \leq M_{BH} < 5.1 \times 10^{6} M_{\odot}$, $5.1
\times 10^{6} M_{\odot} \leq M_{BH} < 1.4 \times 10^{7} M_{\odot}$,
$1.4 \times 10^{7} M_{\odot} \leq M_{BH} < 3.7 \times 10^{7}
M_{\odot}$ or $ \geq 3.7 \times 10^{7} M_{\odot}$ for a given halo
mass at $z=0$. We find that the most massive halos at $z=0$, are most
likely to be descendants of $z=8$ black holes in the mass range between
$10^{6} - 1.4 \times 10^{7} M_{\odot}$ , when compared to the massive
black holes with $M_{BH} > 3.7 \times 10^{7} M_{\odot}$ which is
consistent with the results shown in
Figure~\ref{F:fig_Mhaloz0Mbh}. However, as noted earlier, the
descendants of most massive black holes are relatively massive with
$M_{Halo,z=0} > M_{*,z=0}$ and the probability that halos with masses
less than $10^{13}h^{-1}M_{\odot}$ at $z=0$ are descendants of massive
black holes at $z=8$ is very small.

\section{Conclusions} \label{S:conclusions}

In this paper, we have used the BlueTides MassTracer simulation, a dark
matter-only realization of the BlueTides hydrodynamical simulation 
\citep{2016MNRAS.455.2778F} to trace the host halo descendants of the 
first, rare massive black hole/quasars to the present day.
Due to its large volume and high resolution it is currently 
prohibitive (even on the largest
machines) to run 
 BlueTides (with hydrodynamics and sub-grid physics of galaxy formation)
significantly beyond high redshift.
We match the halos in BlueTides hydrodynamic and the BT MassTracer
simulation which are evolved from the same initial
conditions at $z=8$. The most bound particles in the matched halos of
the dark matter-only simulation are tracked to $z=0$ to identify the
descendants of the most massive black holes.

Our main finding is that the rare quasars with the most massive
black holes do not have the most massive low redshift descendants.
The BTMassTracer simulation is large enough that it contains several
hundred galaxy clusters with masses $\sim 10^{15} M_{\odot}$ at redshift 
$z=0$. These
objects however do not contain these quasar blackholes, which instead
reside in galaxy groups with masses  $\sim 10^{14} M_{\odot}$. We further 
find that for similar range of halo masses at $z=8$, the halo masses of 
the low-redshift descendants of the least massive high-redshift black holes 
at $z=0$ have a distribution 
similar to those of the descendants of most massive high-redshift black holes.
Consistent with our results, \cite{2013MNRAS.436..315F} and \cite{2017arXiv170406050U} find that the most luminous quasars do not live in the most massive halos. Further, \cite{2012ApJ...752...39T} find that the most luminous quasars exist in similar environments when compared with that of their less luminous counterparts. It is also to be noted that \cite{2016Natur.532..340T} report the discovery of a luminous blackhole whose descendant is found in a relatively isolated galaxy.   

We also find that the $z=8$ progenitors of the most massive halos at
$z=0$ do not necessarily host the most massive black holes. We
determined the black hole masses, by identifying the $z=8$ progenitors
of the massive halos at $z=0$ in the BT MassTracer simulation, which
are matched to the $z=8$ halos in the BlueTides hydrodynamic
simulation. The black hole masses of these progenitors are distributed
over a range of masses from $10^{6-8} M_{\odot}$ with the most likely
black hole mass being about $3.2 \times 10^{6} M_{\odot}$. Finally, we
estimate the likelihood that the $z=8$ progenitors will host black
holes of a certain mass, given the halo mass at $z=0$. We find that
for massive halos with masses above $10^{15} h^{-1}M_{\odot}$ at
$z=0$, there is only a $7 \%$ probability of its progenitor hosting a
black hole with $M_{BH} > 3.7 \times 10^{7} M_{\odot}$ at $z=8$. The
black holes are most likely to be in the mass range, $2 \times 10^{6}
- 5 \times 10^{6} M_{\odot}$, where the likelihood is about $40 \%$.

As we have mentioned, our result is not surprising considering observational
evidence that quasars may populate halos which are not the most massive.
It is however counter to the simplest possible scenario, that high peaks 
form high redshift quasars and continue to evolve into the highest peaks at
 $z=0$. This may indicate that abundance matching techniques used in 
large-scale
structure and galaxy property studies may not work well in the context
of quasars. This is consistent with the finding that other environmental 
properties, such as  tidal field strength
appear to be important in the formation of these early supermassive
black holes ( \cite{2017MNRAS.467.4243D}).
\cite{2007ApJ...667...38T} showed using simulations that the material
from the  $10^{6}h^{-1}M_{\odot}$
mass halos hosting the first stars does not end up at later times ($z=6$
in that case) in the halos hosting bright quasars, but instead spans a 
range of environments. Our study can be viewed as an extension of this
work to lower redshift, although we find that galaxy groups are the
most likely host at $z=0$. Observational signatures of this could in principle
include a deficit of old stars in groups at $z=0$ due to strong early
episodes of quasar feedback.

\section*{Acknowledgments}

We acknowledge funding from NSF ACI-1614853, NSF AST-1517593, NSF
AST-1616168 and the BlueWaters PAID program, as well
as NASA ATP grant NNX17AK56G.
 The BT and BTMassTracer simulations were run on the
BlueWaters facility at the National Center for Supercomputing
Applications. We thank Aklant Bhowmick for help with visualizations. 

\bibliographystyle{mnras} \bibliography{draft5}
\end{document}